\documentclass[preprint2]{aastex}

\slugcomment{Astrophys.J. 624 (2005) 898-905}

\shorttitle{Light Curve and Spectrum of SN 2003dh}

\shortauthors{Deng et al.}

\begin{document}

\title{On the Light Curve and Spectrum of SN 2003dh Separated from
the Optical Afterglow of GRB 030329}

\author{J.~Deng\altaffilmark{1,2,3},
 N.~Tominaga\altaffilmark{2},
 P.~A.~Mazzali\altaffilmark{3,4,5},
 K.~Maeda\altaffilmark{2,6}, and
 K.~Nomoto\altaffilmark{2,3}\\
 \vspace{0.8cm}
}

\altaffiltext{1}{National Astronomical Observatories, Chinese
Academy of Sciences, 20A Datun Road, Chaoyang District, Beijing
100012, China.}

\altaffiltext{2}{Department of Astronomy, University of Tokyo,
Hongo 7-3-1, Bunkyo-ku, Tokyo 113-0033, Japan.}

\altaffiltext{3}{Research Center for the Early Universe,
University of Tokyo, Hongo 7-3-1, Bunkyo-ku, Tokyo 113-0033,
Japan.}

\altaffiltext{4}{Max-Planck-Institut f\"ur Astrophysik,
Karl-Schwarzschildstr. 1, D-85748 Garching, Germany.}

\altaffiltext{5}{INAF-Osservatorio Astronomico di Trieste, Via
Tiepolo, 11, I-34131 Trieste, Italy.}

\altaffiltext{6}{Department of Earth Science and Astronomy,
College of Arts and Sciences, University of Tokyo, Komaba 3-8-1,
Meguro-ku, Tokyo 153-8902, Japan.}

\begin{abstract}

The net optical light curves and spectra of the supernova (SN)
2003dh are obtained from the published spectra of GRB 030329,
covering about 6 days before SN maximum to about 60 days after.
The bulk of the $U$-band flux is subtracted from the observed
spectra using early-time afterglow templates, because strong line
blanketing greatly depresses the UV and $U$-band SN flux in a
metal-rich, fast-moving SN atmosphere. The blue-end spectra of the
gamma-ray burst (GRB) connected hypernova SN 1998bw is used to
determine the amount of subtraction. The subtraction of a host
galaxy template affects the late-time results. The derived SN
2003dh light curves are narrower than those of SN 1998bw, rising
as fast before maximum, reaching a possibly fainter maximum, and
then declining $\sim 1.2-1.4$ times faster. We then build $UVOIR$
bolometric SN light curve. Allowing for uncertainties, it can be
reproduced with a spherical ejecta model of $M_{\rm{ej}}\sim 7\pm
3$ $M_\sun$, $E_{K}\sim 3.5\pm 1.5 \times 10^{52}$ ergs, with
$E_{K}/M_{\rm{ej}}\sim 5$ following previous spectrum modeling,
and $M(^{56}{\rm Ni})\sim 0.4_{-0.1}^{+0.15}$ $M_\sun$. This
suggests a progenitor main-sequence mass of $\sim 25-40$ $M_\sun$,
lower than SN 1998bw but significantly higher than normal Type Ic
SNe and the GRB-unrelated hypernova SN 2002ap.

\end{abstract}


\section{INTRODUCTION}

Supernova SN 2003dh was discovered in the spectra of the optical
transient (OT) of the nearby gamma-ray burst GRB 030329
($z=0.1685$; e.g., \citealt{sta03b,hjo03}), providing a strong
confirmation that long GRBs originate from core collapse of very
massive stars (e.g., the ``collapsar'' model; \citealt{mac99}).
Previous clear evidence of the GRB-SN connection, although not as
direct, was the temporal and spatial coincidence between SN 1998bw
and another nearby GRB 980425 ($z=0.0085$; \citealt{gal98}). SN
1998bw displayed unusually broad spectral features that require a
high explosion energy to explain, more than 10 times that of an
ordinary SN \citep{iwa98,woo99}, and because of this it was termed
a hypernova \citep{nom04}. The spectral features of GRB 030329/SN
2003dh have also been shown to be hypernova-like
\citep{sta03b,kaw03,hjo03,mat03,kos04}.

Various efforts have been made to separate the SN light from the
underlying optical afterglow (OA) of GRB 030329, yielding
different results (see \citealt{lip04} and references therein).
The difficulty lies in the fact that the brightness of the OA
thoroughly eclipsed that of the SN for the first week and kept
competing with it for at least 1 month. This is unlike the case of
GRB 980425/SN 1998bw, in which an OA was not detected. The problem
is further compounded by the contribution of the host galaxy to
any spectroscopy and photometry.

\citet{mat03} and \citet{hjo03} used a least-squares fit to
decompose their observed spectra of the first month. They adopted
the spectra of SN 1998bw and other hypernovae as templates and
assumed an OA template using either an early spectrum or a
power-law continuum. The similarities of the first-month spectra
to those of SN 1998bw were revealed. However, the assumption
implied in this method, that each spectrum of SN 2003dh is
identical to one or the other of the known hypernovae, can not be
guaranteed a priori. \citet{mat03} found that the light curve (LC)
of SN 2003dh also follows that of SN 1998bw (but is fainter by
$\sim 0.2$ mag), comparing their derived $R_{\rm C}$-band light
curve (LC) to the $V$-band LC of SN 1998bw without calculating
real $K$-corrections. \citet{hjo03}, on the other hand, obtained a
$V$-band LC with a peak $\sim 5$ days earlier, slightly brighter
at maximum, and declining $\sim 1.4$ times faster after maximum
than SN 1998bw.

\citet{kos04} subtracted a host galaxy template from their
$\sim$day 40 and $\sim$day 80 spectra, assuming a negligible OA
contribution as extrapolated from \citet{mat03}. Their SN spectra
in the red part seem more similar to another hypernova, SN 1997ef
\citep{maz00} than to SN 1998bw (see also \citealt{kaw03,maz03}).
They found that the decline rate of SN brightness between these
two epochs is similar to that of SN 1998bw.

\citet{blo04} and \citet{lip04} derived the SN LC directly from
the OT photometry, without resorting to spectra. \citet{blo04}
assumed a SN 1998bw-like LC shape and suggested that SN 2003dh is
1.5 times brighter than SN 1998bw, after correcting for an assumed
host galaxy extinction of $A(V)=0.3$ mag. However, their
photometry was limited to the first 23 days, and was well sampled
only until day 12, when the SN component was still weak relative
to the OA, and hence was unable to constrain the SN LC well.
\citet{lip04}, on the other hand, analyzed a dense $BVRI$ data set
of the OT photometry extending to day 80. They constructed LCs for
SN 2003dh by time-stretching those of SN 1998bw by a factor of 0.8
and lowering the brightness by 0.3 mag. They implicitly assumed
that the color evolution also follows that of SN 1998bw,
time-stretched by the same factor.

To address this problem with the view of theoretical modeling,
\citet{maz03} pointed out that a reasonable spectrum of a Type Ic
SN or hypernovae must show a flux deficiency to the blue of $\sim
3600$ \AA, like SNe Ia, owing to strong blanketing effects of
dense metal lines (e.g., \citealt{bar99,maz00}). They attributed
the blue flux of the OT spectrum to the OA and used this to pivot
the subtraction to obtain the SN spectrum. However, only three
spectra at typical epochs were processed and modelled in that
paper. The three bolometric points so derived were nonetheless
shown to be consistent with a synthetic LC based on spectral
models, which peaks $\sim 5$ days earlier, has a maximum
brightness $\sim 0.35$ mag fainter, and at around 1 month is
fainter by $\sim 0.6$ mag than SN 1998bw.

It is important that the spectra and LCs of SN 2003dh are
correctly derived, specifying the similarities and differences to
SN 1998bw and to other Type Ic SNe and hypernovae, whether
GRB-related or not. Recently, another nearby GRB-SN association
was discovered, i.e., GRB 031203/SN 2003lw ($z=0.1055$; e.g.,
\citealt{tom04,mal04}). Interestingly, SN 2003lw was also claimed
to be similar to SN 1998bw, although the observations were
dominated by the light from the host galaxy, and the SN suffered
from serious extinction. Other observers emphasized the
differences with respect to SN 1998bw \citep{cob04,gal04}. Other
cases of possible SNe in GRBs have been reported, virtually all of
them based on the detection of a SN 1998bw-like ``bump'' in the OT
LC (e.g., \citealt{blo02,gar03}). However, GRB 031211/SN 2002lt
may be an exception ($z=1.006$; \citealt{del03}), since it has
been suggested that it resembles the ordinary Type Ic SN 1994I
(e.g., \citealt{bar99}) or the weak version of hypernovae, SN
2002ap (e.g., \citealt{maz02}).

On the other hand, for studies on the late evolution of the OA of
GRB 030329, it is also crucial to consider how the OA and SN
components are separated, as shown by \citet{lip04}.

In this paper, we update the spectra and LCs of SN 2003dh by
applying the decomposition method of \citet{maz03} on all the
observed spectra available to us. However, unlike \citet{maz03},
who for simplicity removed as much blue flux as possible from the
OT spectra, we also refer to the blue part of SN 1998bw spectra
that begins to show the flux deficiency to determine the amount of
OA subtraction. The bolometric LCs resulting from the two
approaches are compared to one another and to model LCs.

\section{SPECTRUM DECOMPOSITION}

We studied 14 observed spectra published by \citet{mat03}, those
taken with the MMT, Magellan, and Keck telescopes, spanning from
April 1 to May 24 (see also \citealt{sta03b}), and the May 8/9 and
June 22 spectra taken with the Subaru telescope
\citep{kaw03,kos04}. The $B$-, $V$-, and $R_{\rm C}$-band
magnitudes that we derived from each \citet{mat03} spectrum,
except that of May 24, are consistent to within $\sim\pm 0.15$ mag
with the photometry of \citet{lip04} and with that reported in
GCNs. The differences in colors are smaller, i.e., $\lesssim 0.1$
mag. The May 24 spectrum was of rather poor quality and was taken
during the ``Jitter Episode'' when the OT varied rapidly by $\sim
0.5$ mag on a timescale of a few days \citep{sta03a,mat03}. It was
fainter by $\sim 1$ mag in the $R_{\rm C}$-band than the above
photometry, which had an almost daily coverage of the Jitter
Episode. We re-calibrated its flux using $R_{\rm C}=21.4\pm 0.15$
from the photometry. To compensate for similar flux discrepancies,
the May and June Subaru spectra were calibrated to $R_{\rm
C}=20.9\pm 0.1$ and $21.8\pm 0.1$, respectively, following
\citet{kaw03} and \citet{kos04}.

All the spectra were converted to the rest frame of the GRB/SN
($z=0.1685$), adopting a distance modulus of 39.54. We adopted the
spectrum of NGC 3125, which was found by \citet{kos04} to show
good coincidence to the observed narrow-line ratios, as the host
galaxy template, and calibrated its flux to $V=22.7\pm 0.3$
\citep{fru03}. This host galaxy template was subtracted from the
observations, whose flux contribution steadily increases from
$<10$\% in the first 20 days to $\sim 40\%-60\%$ around 60 days
(see Figure \ref{fig1}), to derive the OT spectra. This operation,
however, did not fully eliminate the narrow emissions, so the
residuals, artifacts, and telluric lines were removed by hand.

We used each of the April 1, 2, 3, and 4 OT spectra as the OA
template. The April 1, 2, and 3 spectra, after sufficient
smoothing, can be well fitted using a power law $f_\lambda\propto
\lambda^{-\beta}$, with $\beta=-1.06\pm 0.00$, $-1.12\pm 0.00$,
and $-0.97\pm 0.00$, respectively. The April 4 spectrum, however,
shows significant deviations from a power-law, which are not like
SN features (see below). These fitting power-laws and the smoothed
April 4 spectrum were used in our actual spectrum subtraction.

As stated above, we scaled the OA template in flux and subtracted
it from the OT spectrum to obtain the SN component, taking into
account the fact that strong line blanketing should have greatly
depressed the SN flux to the blue of $\sim 3600$ \AA. Line
blanketing (or blocking) describes the significant strengthening
of UV/$U$ continuum opacity caused by absorptions of numerous
overlapping metal lines, mostly of Fe-group elements and Si, in a
metal-rich, fast-moving SN atmosphere such as that of a SN Ia
(e.g., \citealt{pau96}), or of a SN Ic (e.g., \citealt{bar99}), or
of a hypernova (e.g., \citealt{maz00}).

We derived a quantitative criterion for the spectrum subtraction
from the spectra of SN 1998bw \citep{pat01}, which have good
$U$-band coverage. In the SN 1998bw spectra, the flux drops
rapidly from $\sim 4000$ to $\sim 3600$ \AA, along the \ion{Ca}{2}
H\&K line profile, to reach the flux-deficiency range. The flux
ratio of two reference wavelengths,
$f_{4000\rm\AA}/f_{3350\rm\AA}$, evolves from $\sim 2$ in the
first 15 days to $\sim 3$ between 15 and 40 days, and to $\sim 4$
in the rest of the photospheric epoch. We imposed these values on
the derived SN 2003dh spectra (case I). On the other hand, we also
tried the simple treatment of \citet{maz03}, forcing the minimal
SN spectral flux in the $U$-band to zero (case II). Since the OT
spectra were noisy, in order to apply these criteria, we smoothed
them before subtraction by averaging every 100 or 75 wavelengths
points.

To determine the OA flux for epochs of the Subaru spectra that
have insufficient blue coverage, we fitted the OA flux evolution
derived from decomposing the earlier spectra with a power law
$f\propto t^{-\alpha}$. The best-fit power-law index is about
$2.75\pm 0.15$ for case I and about $2.40\pm 0.15$ for case II.

We show in Figure \ref{fig2} the spectra of SN 2003dh between
April 6 and June 22 that we derived from the smoothed OT spectra
in case I by subtracting the April 3 OA template, except for April
7. Using the different OA templates mentioned above did not change
the overall spectral appearance in most cases and affects the
continuum shape, and hence the multiband photometry, only
slightly. However, for April 7, if any of the above OA templates
were used, the derived spectrum turned out to be too blue to look
like one of any Type Ic SN or hypernova. We assumed an OA spectrum
of $f_\lambda\propto \lambda^{-1.5}$ to get the ``reasonable''
April 7 SN spectrum shown in Figure \ref{fig2}. The above problem
of the April 7 spectrum was likely caused by observational
uncertainties, because the difference between $B-R_{\rm C}$ of
this OA spectrum and that of the April 3 spectrum, i.e., $\sim
0.23$, is the same as that between the observed April 7 spectrum
and the \citet{lip04} photometry.

The resulting SN spectra in case II are similar to those of case
I, but they are fainter and redder since more OA flux was
subtracted especially in the $UB$ bands. To exemplify the
difference between these two cases, in Figure \ref{fig3} we
compare the derived April 8 SN spectra with the SN 1998bw spectrum
at a similar epoch. The SN spectrum in case I matches both the red
and blue parts of the SN 1998bw spectrum well, while that in case
II shows a significant blue excess relative to the latter. The
difference becomes small for the spectra of May and later, when
the OA has greatly faded and contributes about 10\% -- 15\% of the
total flux or less (see Figure \ref{fig1}).

The April SN spectra resemble those of SN 1998bw at similar
epochs, as discovered by \citet{sta03b}, \citet{hjo03}, and
\citet{mat03}. On the other hand, \citet{kaw03} and \citet{kos04}
pointed out that the red part of the May and June Subaru spectra
resembles the SN 1997ef spectra, where the \ion{O}{1} $\lambda
7774$ and \ion{Ca}{2} IR triplet lines are well separated, more
than they do the SN 1998bw spectra, where these two lines still
strongly blend at such epochs (see also \citealt{maz03}). This
seems to be confirmed by the May 24 spectrum of \citet{mat03}, as
shown in Figure {\ref{fig4}}, where we compare the derived SN
spectrum with those of SN 1997ef and SN 1998bw at similar epochs.
Although the spectrum is noisy, it seems to show separate
\ion{O}{1} $\lambda 7774$ and \ion{Ca}{2} IR triplet lines,
similar to SN 1997ef and unlike SN 1998bw (see Figure \ref{fig2}
for a smoothed spectrum), if the red end has not been much
polluted by the second-order spectrum.

We cannot find convincing SN features in the derived spectra on
April 4 and 5, which are also shown in Figure \ref{fig2} for
comparison. (The April 4 spectrum shown was obtained using the
April 2 OA template) First of all, they only contribute a few
percent to the total flux, which is within the flux calibration
error in the observations (K. Z. Stanek \& T. Matheson 2004,
private communication; see also \citealt{mat03}). In addition, the
derived April 4 spectrum is continuum-like, similar to the derived
April 3 spectrum in \citet{hjo03}, while the hypernova SN 2002ap
at $\sim 2-4$ days already displays clear, broad line features
\citep{maz02,kin02}. Furthermore, even if the dubious ``broad
features'' in the derived April 5 spectrum are real, they do not
match the SN features on April 6 and later, and hence are unlikely
to be related to the SN. So we confirm that the April 3 -- 5 OT
``color event'' \citep{mat03}, corresponding to the LC ``bump D''
of \citet{lip04} and close to the supposed jet break, is not of SN
origin.

\section{LIGHT CURVES}

We assembled the rest-frame $UBVR_{\rm C}I_{\rm C}$ multicolor LCs
for SN 2003dh from broadband spectroscopy of the derived SN
spectra, excluding April 7 (see above), and corrected for the
Galactic reddening of $E(B-V)=0.025$ \citep{sch98}. The LCs in
both case I and II are shown in Figure \ref{fig5} and compared
with those of SN 1998bw [$z=0.0085$, $\mu=32.89$, and $A(V)=0.2$;
\citealt{gal98,pat01}]. These are the results averaged over the
cases of different OA templates. We list in Table 1 the LC data in
case I and their error bars, which vary between $\pm 0.2$ and $\pm
0.4$ mag. We estimated the error bars from statistical
uncertainty; flux calibration uncertainty of the observed spectra,
i.e., $\pm 0.15$ mag for the Matheson spectra and $\pm 0.1$ mag
for the Subaru spectra; and flux calibration uncertainty of the
host galaxy, i.e., $\pm 0.3$ mag. They are not shown in Figure
\ref{fig5}, for clarity.

The derived SN~2003dh LCs are more similar to those of SN 1998bw
than to those of any other known hypernova or Type Ic SN, but on
the other hand, they are also narrower than those of SN 1998bw in
both case I and II. The case I LCs are as bright as SN 1998bw
before maximum but fainter by about 0.3 -- 0.6 mag after maximum.
Those in case II are even fainter, particularly in the $U$ and $B$
bands. We think that case II has underestimated the $UB$ flux to
some extent, by comparison with SN 1998bw and other hypernovae and
Type Ic SNe. Case I seems therefore more realistic.

Our case I results are consistent with the apparent $V$-band LC
derived by \citet{hjo03} through decomposing the VLT spectra of
the first month. Their best-fit OA spectrum,
$f_\lambda\propto\lambda^{-0.8}$, is not as steep as our
templates, however. For comparison, we calculated the apparent $V$
magnitudes of our case I SN spectra converted to the observer
frame. Their magnitudes on rest-frame day 28, 20, and 8 are only
fainter than ours by $\lesssim 0.1$ mag, while their day 10 is
$\sim 0.2$ mag brighter. The differences are within our respective
error bars. \citet{hjo03} also showed a day 4 magnitude much
brighter than SN 1998bw. However, the corresponding spectrum is
not SN-like. As argued by \citet{mat03} and confirmed by us, SN
features did not emerge until April 6. It would be useful to apply
the same treatment to the VLT spectra. For example, the LC around
maximum could be defined much better using the VLT spectrum on
rest-frame day 16.

The real SN LC inferred from the spectrum decomposition of
\citet{mat03} may also be narrower than that of SN 1998bw, similar
to our results, although it was stated by those authors that the
two SNe closely resemble each other. They were able to model the
observed $R_{\rm C}$-band LC by combining the derived OA $R_{\rm
C}$-band LC and a $V$-band LC of SN 1998bw, corrected for time
dilation. However, as pointed out by \citet{lip04}, they would
have suggested a SN LC $\sim 0.4-0.6$ mag fainter than SN 1998bw
if $K$-corrections had been taken into account. $K$-corrections
for SN 1998bw have been calculated for various redshifts (J. Deng
2005, in preparation). The one of interest here, $R_{\rm
C}(z=0.1685)-V(z=0.0085)$, varies slightly between $-0.29$ and
$-0.34$ in the first 100 days. The redshifted and $K$-corrected
$V$-band LC of SN 1998bw exceeds the observed $R_{\rm C}$-band
brightness of the GRB/SN at some epochs of the Jitter Episode
($\sim$ day 50 -- 70), confirming the findings of \citet{lip04}.
On the other hand, upon close inspection, the combined LC shown in
Figure 13 of \citet{mat03} seems to underestimate the observations
at the early epochs, i.e., before SN maximum.

The SN LCs recommended by \citet{lip04}, i.e., those of SN 1998bw
stretched by 0.8 times and attenuated by 0.3 mag, are $\sim
0.05-0.55$ mag fainter, in the rest frame, than our case I
results, depending on the individual epochs and the bands. Our
$B$- and $U$-band LCs also decline faster after day 30 than the
Lipkin LCs in both case I and case II. Only the $V$- and $R_{\rm
C}$-band LCs in case II seem to match the Lipkin ones.
\citet{lip04} argued that the $\Delta m=0.3$ mag is the lower
limit on their attenuation to meet the brightness minimum in the
Jitter Episode. Our results obviously also meet that requirement.

We built the $UVOIR$ bolometric LCs for SN 2003dh using the above
multicolor LCs in order to compare with theoretical models.
However, we did not follow the common practice in which a spectral
energy distribution (SED) was constructed and integrated, because
of the lack of the rest frame $I_{\rm C}$-band LC and near-IR
data. Instead, we calculated the bolometric corrections (BCs) of
SN 1998bw in each band and applied them to the SN 2003dh
multicolor LCs (see Table 1). This is justified by the overall
spectral similarity of these two SNe. The $R_{\rm C}$-band BC of
SN 1997ef seems more suitable for May and June, because the
spectra in this period are more similar in the red part to SN
1997ef than to 1998bw. We derived the $R_{\rm C}$-band BC in such
epochs from the SN 1997ef spectra \citep{maz00} but found that its
value, $\sim 0.5 - 0.6$, is similar to that of SN 1998bw, $\sim
0.6$. Therefore, we ignored the difference between them.

We averaged the bolometric LCs converted from the individual LC of
each band to make up for the color difference between SN~2003dh
and SN 1998bw. The average was taken over the $B$ and $V$ bands
for most epochs, or over the $B$, $V$, and $R_{\rm C}$ bands, if
the last was available. We excluded the $U$-band to avoid
potentially large uncertainties. The bolometric LC converted from
the $U$-band LC is more or less consistent with the average one in
case I, but that in case II is $\sim 0.4-0.5$ mag fainter than the
average one. In many cases, the $U$ band contributes to only a
small fraction of the total flux of Type Ic SNe, owing to line
blanketing, so the bolometric LC may not be well correlated with
the $U$-band LC. Moreover, the BCs of SN 1998bw in the $U$ band
may not be as reliable as in the other bands because of the common
difficulties related to $U$-band photometry. In contrast, the
bolometric LCs converted from the $B$-, $V$-, and $R_{\rm C}$-band
LCs agree with the average one within $\sim\pm 0.15$ mag in both
case I and case II.

We show our average bolometric LCs of both case I and case II in
Figure \ref{fig6} ({\em top}), where SN 2003dh is compared with
the hypernovae SN 1998bw \citep{pat01}, SN 1997ef \citep{maz00},
and SN 2002ap \citep{maz02,yos03} and normal Type Ic SN 1994I (B.
P. Schmidt \& R. P. Kirshner 1993, private communication;
\citealt{nom94,ric96}). Although as bright as SN 1998bw before
maximum, SN 2003dh was significantly fainter after maximum,
resulting in relatively narrow LCs. In case I, the brightness
discrepancy increases from $\sim 0.35$ mag around rest-frame day
20 to $\sim 0.55$ mag around day 70, while in case II it is mostly
$\sim 0.5 - 0.6$ mag after maximum. The LC decline is about 1.4
times faster than SN 1998bw in case I, confirming the result of
\citet{hjo03}. The case II LC resembles the solution suggested by
\citet{lip04}, i.e., that the LC of SN 2003dh is similar to that
of SN 1998bw stretched by 0.8 times and attenuated by 0.3 mag. On
the other hand, the above differences notwithstanding, SN 2003dh
is still much more similar to SN 1998bw than to the other SNe
shown in the figure with regard to the bolometric LC.

We suggest that the peak brightness of SN 2003dh is lower than
that of SN 1998bw, even though our LCs do not capture the exact
time of peak, which lies between rest-frame day 10 and 15
according to \citet{hjo03}. For case II this can easily be seen in
Figure \ref{fig6}. However, in case I, our April 8 and 10
magnitude points lie above the bolometric LC of SN 1998bw and
could lead to a peak almost as bright as SN 1998bw. On the other
hand, the corresponding spectra at these epochs in Figure
\ref{fig2} display SN features not as prominent as on April 6 and
9. So it is possible that there is still some residual OA
continuum, whose spectrum would not be a power law. The actual SN
LC may pass between the bright April 8 and 10 points and the faint
April 6 and 9 points, leading to a somewhat fainter peak than SN
1998bw.

We modeled the bolometric LCs starting from the ejecta model COMDH
constructed in \citet{maz03}. We modified its mass $M_{\rm{ej}}$,
kinetic energy $E_{K}$, and ejected $^{56}$Ni mass $M(^{56}{\rm
Ni})$, while keeping the ratio $E_{K}/M_{\rm{ej}}$ constant. COMDH
is a one-dimensional spherical density structure against velocity
that was derived from theoretically synthesizing the SN spectra of
April 10, 24, and May 8/9. It has $M_{\rm{ej}}\sim 8$ $M_\sun$,
$E_{K}\sim 4\times 10^{52}$ ergs, and a $^{56}$Ni mass of 0.35
$M_\sun$. The density structure above 25,000 $\rm{km~s^{-1}}$
mimics the hypernova model of SN 1998bw \citep{nak01} to reproduce
the SN 1998bw-like early-time spectra and very broad lines, while
below 15,000 $\rm{km~s^{-1}}$ it is adjusted so as to follow the
spectral evolution into the SN 1997ef-like later epoch. Our
derived SN spectra are similar to those studied by \citet{maz03}
as far as spectral features are concerned, and the small
differences in the total flux are not expected to modify the main
results of spectral modeling obtained in that paper. Therefore, we
did not model the spectra again, but rather we fixed
$E_{K}/M_{\rm{ej}}$ at the original COMDH value, i.e., $\sim 5$,
because this ratio, which characterizes the line widths and the
strength of line-blending, is the key parameter in spectrum
modeling.

We used the same one-dimensional SN radiation hydrodynamical and
gamma-ray transfer codes \citep{iwa00} used in \citet{maz03}, but
we adopted the treatment of the Eddington factor proposed by
\citet{gom96}, instead of the simple Eddington approximation, and
we approximated the Rosseland mean opacity using an empirical
relation to the electron-scattering opacity derived from the TOPS
database \citep{mag95}, rather than assuming a constant line
opacity. Isotope $^{56}$Ni was assumed to be homogeneously mixed
throughout the ejecta in order to reproduce the rapid brightening
before maximum, as in other hypernovae, such as SN 1998bw
\citep{nak01}, SN 1997ef \citep{iwa00}, and SN 2002ap
\citep{maz02}. Since COMDH is not the result of an explosion
simulation, the other composition was assumed to be 90\% O and
10\% Si. Our synthetic LC of the original COMDH is consistent with
that of \citet{maz03}.

Our bolometric LC in case I is best-fitted with the slightly
reduced values $M_{\rm{ej}}\sim 7$ $M_\sun$, $E_{K}\sim 3.5\times
10^{52}$ ergs, and $M(^{56}{\rm Ni})\sim 0.4$ $M_\sun$, while that
in case II can be well reproduced using the original COMDH. The
best-fitting model LCs are shown in Figure \ref{fig6} ({\em
bottom}).

The fitting model parameters for our preferred case I can be
relaxed to $M_{\rm{ej}}\sim 7\pm 3$ $M_\sun$ and $E_{K}\sim 3.5\pm
1.5 \times 10^{52}$ ergs, with $E_{K}/M_{\rm{ej}} \sim 5$, if the
error bars shown in Figure \ref{fig6} are taken into
consideration. To derive the upper and lower limits, we assumed
that we had either under- or overestimated the SN flux throughout
the postmaximum epochs. We did not place much weight on fitting
the premaximum LC, because it is not well defined, and because the
model LC in that phase is more sensitive to the $^{56}$Ni
distribution. Finally, we constrained the most likely range for
the $^{56}$Ni mass to $M(^{56}{\rm Ni})\sim 0.4_{-0.1}^{+0.15}$
$M_\sun$.

\section{DISCUSSION}

We discussed an optimal OA subtraction method to separate the SN
spectra from the OA of GRB 030329 and to construct the $UVOIR$
bolometric LC of SN 2003dh. The LC is narrower than that of SN
1998bw, rising as fast before maximum but reaching a somewhat
fainter maximum and declining about 1.2 -- 1.4 times faster
afterward. Our spherical LC model parameters, $M_{\rm{ej}}\sim
7\pm 3$ $M_\sun$ and $E_{K}\sim 3.5\pm 1.5 \times 10^{52}$ ergs
($E_{K}/M_{\rm{ej}}\sim 5$), allowing for LC uncertainties, are
lower than SN 1998bw, which is another GRB-associated hypernova
and has been modelled with $M_{\rm{ej}}\sim 10$ $M_\sun$ and
$E_{K}\sim 4 - 5 \times 10^{52}$ ergs \citep{nak01,mae03}. SN
1997ef, a hypernova displaying no GRB, as well as it replica SN
1997dq \citep{maz04}, is as massive as SN 1998bw and SN 2003dh but
less energetic ($E_{K}\sim 1 - 2 \times 10^{52}$ ergs;
\citealt{iwa00,maz00}).

We suggest that SN 2003dh/GRB 030329 is the explosion of an $\sim
6-14$ $M_\sun$ C+O star, assuming a central remnant of $\sim 2-4$
$M_\sun$, which may evolve from an $\sim 25-40$ $M_\sun$
main-sequence star \citep{nom88} losing its H and He envelopes
through stellar winds or binary accretion \citep{nom95}. The upper
limit is comparable to the values used for SN 1998bw, and the
lower limit is well above that for SN 2002ap \citep{maz02}, a
marginal hypernova without an associated GRB. The parameter range
could be further narrowed by spectral modeling, as done in
\citet{maz03}. However, the result in that paper could be updated
if the spectra presented here, which are a little brighter and
extend to later epochs, are modelled, although the characteristic
parameter $E_{K}/M_{\rm{ej}}$ is expected to have a similar value.

The model parameters will probably be modified when asymmetry is
considered, as was the case for SN 1998bw (\citealt{hof99,mae02}).
The changes in $M_{\rm{ej}}$ and $^{56}$Ni mass are not expected
to be large, while $E_{K}$ can be significantly reduced in an
asymmetric model, depending on the geometry. In the case of SN
1998bw, \citet{mae02} found an axial ratio of about 3:2 and a
viewing angle of about $15^{\circ}$ from the pole, by modeling the
late-time spectra. Their best-fitting $E_{K}$ is $\sim 10^{52}$
ergs, i.e., one-fourth that in the spherical model of
\citet{nak01}. The SN 2003dh asymmetry is probably similar. This
is modest compared to the GRB. \citet{kaw03} detected marginal
polarization ($<1\%$) in 2003 May and suggested that it was mostly
of local interstellar origin. \citet{gre03} reported an OT
polarization level of $\sim 0.5\%-2\%$ at times between $\sim 10$
and 40 days after the GRB and assumed that interstellar
polarization was negligible. This polarization level seems
consistent with other envelope-stripped core-collapse SNe (e.g.,
\citealt{wan01,leo02}) and may suggest an asphericity of $\sim
15\%-30\%$ \citep{hof91}.

The evolution of the OA flux after April 3 -- 5 derived from our
case I spectrum decomposition can be fitted using a power-law
decay with an index of $\alpha\sim 2.7$ (see Figure \ref{fig7} for
an example). This is steeper than what was suggested by
\citet{lip04}, i.e., $\alpha\sim 2.0$ after the jet break that
occurred around day 3 -- 8. On the other hand, we can also fit our
results using a broken power law, i.e., $\alpha\sim 2.0$ before
$\sim$ day 10 -- 11 and $\alpha\sim 2.9$ after that, in view of
the two-jet model of \citet{ber03}, who invoked a second, wide jet
to explain radio observations and suggested that its break took
place around day 10. However, the above indices are statistically
rough due to the small number of data points, and we cannot tell
which fitting is better.

We have corrected for the Galactic extinction only. \citet{mat03}
derived a local reddening of $E(B-V)\sim 0.03-0.09$ from the
Balmer decrement of the host galaxy. However, the assumption of
Case B recombination underlying this estimate may not be correct.
Moreover, their early OT SED constructed from optical-IR
photometry is consistent with the case of no local extinction. The
negligible local interstellar polarization ($<0.3\%$) argued by
\citet{gre03} corresponds to a negligible local $E(B-V)$ $< 0.03$
mag \citep{ser75}, while the values discussed by \citet{kaw03} may
suggest a local $E(B-V)< 0.06-0.1$ mag. \citet{blo04} obtained an
upper limit for the local extinction, i.e., $A(V)< 0.3$ mag, from
their early optical-IR photometry, although they adopted this
upper limit as the real local extinction value and applied it to
their SN LC. \v{S}imon, Hudec, \& Pizzichini (2004) also argued
for $E(B-V)< 0.1$ mag, comparing the colors of the OA with a group
of 25 OAs.

It is interesting that the three SNe with the most convincing GRB
connections, i.e., SN 1998bw, SN 2003dh, and SN 2003lw, happen to
have relatively similar spectra and LCs, considering the large
differences between their respective GRBs and the general
heterogeneity among hypernovae and among Type Ic SNe. Studies of
SN 2003lw revealed SN 1998bw-like spectra and LCs that are
probably $\sim 0.3-0.7$ mag brighter and $\sim 1.1-1.2$ times
broader than SN 1998bw (\citealt{mal04}; P. A. Mazzali 2005, in
preparation), suggesting a somewhat more massive star. GRB 980425
and GRB 031203 were both extremely weak gamma-ray events,
$E_\gamma^{\rm iso}$ being $\sim 8\times 10^{47}$ ergs for GRB
980425 \citep{gal98} and $\lesssim 10^{50}$ ergs for GRB 031203
\citep{saz04}, and their respective OA was not detected or was
detected only marginally detected \citep{mal04}. In contrast,
$E_\gamma^{\rm iso}$ of GRB 030329 is as high as $\sim 1\times
10^{52}$ ergs \citep{pri03} and its OA dominated the early optical
observations, although the beaming-corrected gamma-ray energy,
$\sim 5\times 10^{49}$ ergs, is still significantly lower than the
bulk of normal GRBs \citep{ber03}.

Does this imply any correlation between the progenitor mass, the
SN kinetic energy, and the GRB/OA intensity? On the one hand, it
is possible that a more massive star causes the deposition of more
energy near the central explosion engine, e.g., the initial jet.
On the other hand, one may speculate that the GRB/OA jet has
dissipated more energy to the SN ejecta after penetrating the more
massive star, and meanwhile may also have given out more energy to
some less relativistic jet component, such as the wide radio
component proposed by \citet{ber03}. Note that the bulk of the
explosion energy for all three events lies in the SN kinetic
energy $E_{\rm K}$, not in the GRB and its afterglow. Thus a small
relative change in $E_{\rm K}$, e.g., a positive one, can
accompany a big change in the GRB energy, positive or not. A
positive correlation between $E_{\rm K}$ and SN ejecta mass can
help prevent big differences in the SN spectra, characterized by
$E_{\rm K}/M_{\rm ej}$, and LCs, which have a typical time-scale
of $\sim M_{\rm ej}^{3/4}/E_{\rm K}^{1/4}$ \citep{arn82}. The
possibility of significantly different viewing angles for the
three GRBs seems to be disfavored by the spectral similarity of
the related SNe unless the SN asphericity is small. \citet{ram04}
suggest that at least GRB 030329 and GRB 031203 are compatible
with the premise that they are the same event viewed very close to
the jet axis and a few degrees away from it, respectively. The
viewing angle of GRB 980425 was determined by \citet{mae02} to be
$\sim 15^{\circ}$, on the basis of the nebular line profiles and
of aspherical explosion models. This may require an intrinsically
weaker GRB.

More examples of GRB-SN connections are required to clarify this
problem. \citet{del03} argued that the SN 2002lt hosted in GRB
021211 is similar to the normal Type Ic SN 1994I. However, the
available rest-frame $U$-band photometry does not exclude a SN LC
as bright and as broad as SN 1998bw. In the case of GRB 011121,
\citet{gar03} detected a LC ``bump'' in the OT. Based on
photometry, they argued for a very blue SN (named SN 2001ke),
different from SN 1998bw both in colors and in temporal evolution.
The only available spectrum is unfortunately too noisy to show
obvious SN features.

\acknowledgments
 We thank T. Matheson, K. S. Kawabata, and G.
Kosugi for the spectra of GRB 030329/SN 2003dh; F. Patat for the
spectra of SN 1998bw; and K. Z. Stanek for important and very
helpful comments on the manuscript.

\figcaption[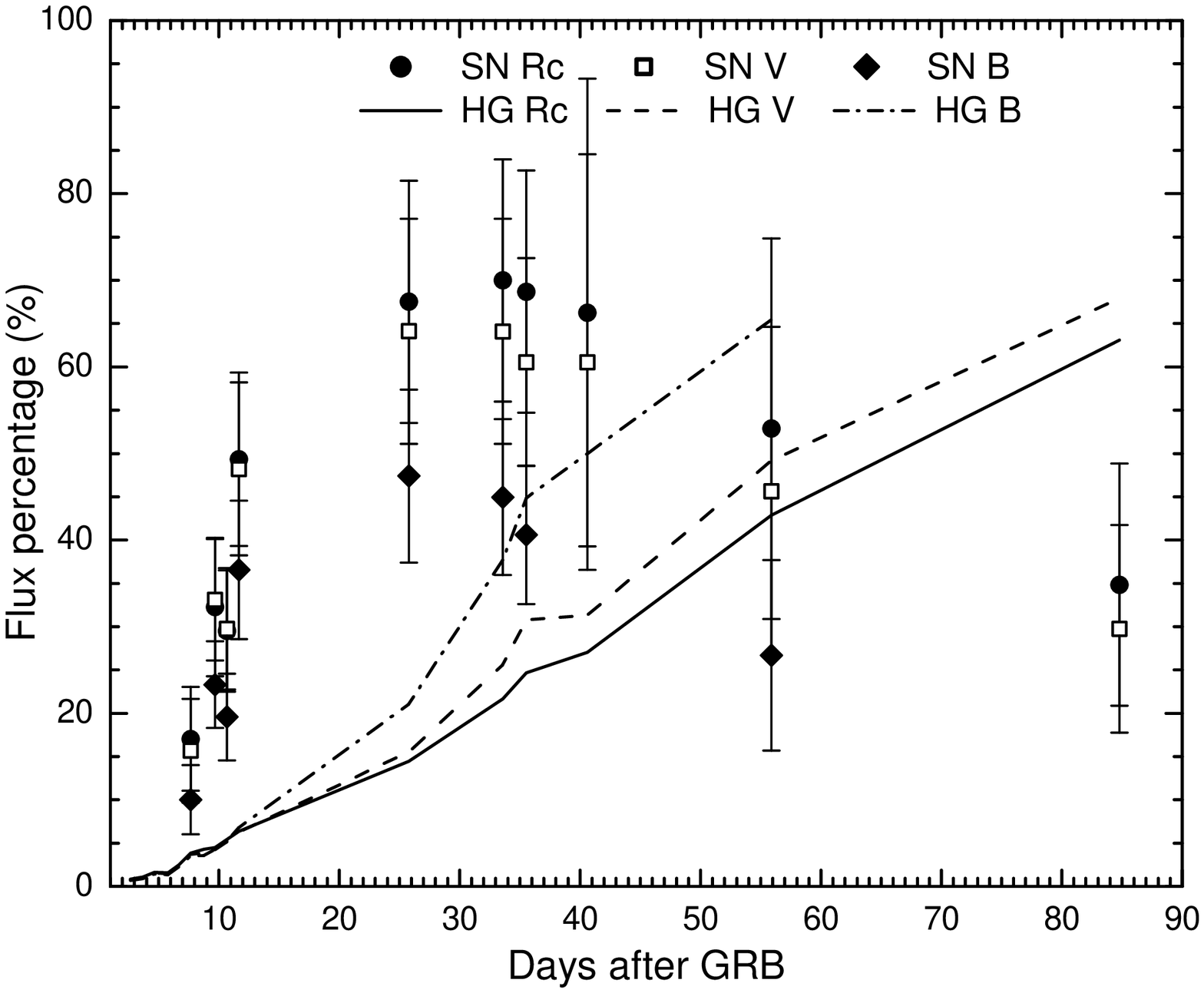]{Contribution of the derived SN 2003dh spectrum
in case I to the OT of GRB 030329 in the observer-frame $R_{\rm
C}$ ({\em circles}), $V$ ({\em squares}), and $B$ ({\em diamonds})
bands, and that of the host galaxy template in these bands ({\em
solid}, {\em dashed}, and {\em dash-dotted line}, respectively).
Results shown are averaged over the cases of different OA spectrum
templates. Error bars are estimated from statistical uncertainty
amd flux calibration uncertainty of the observed spectra and of
the host galaxy. \label{fig1}}

\figcaption[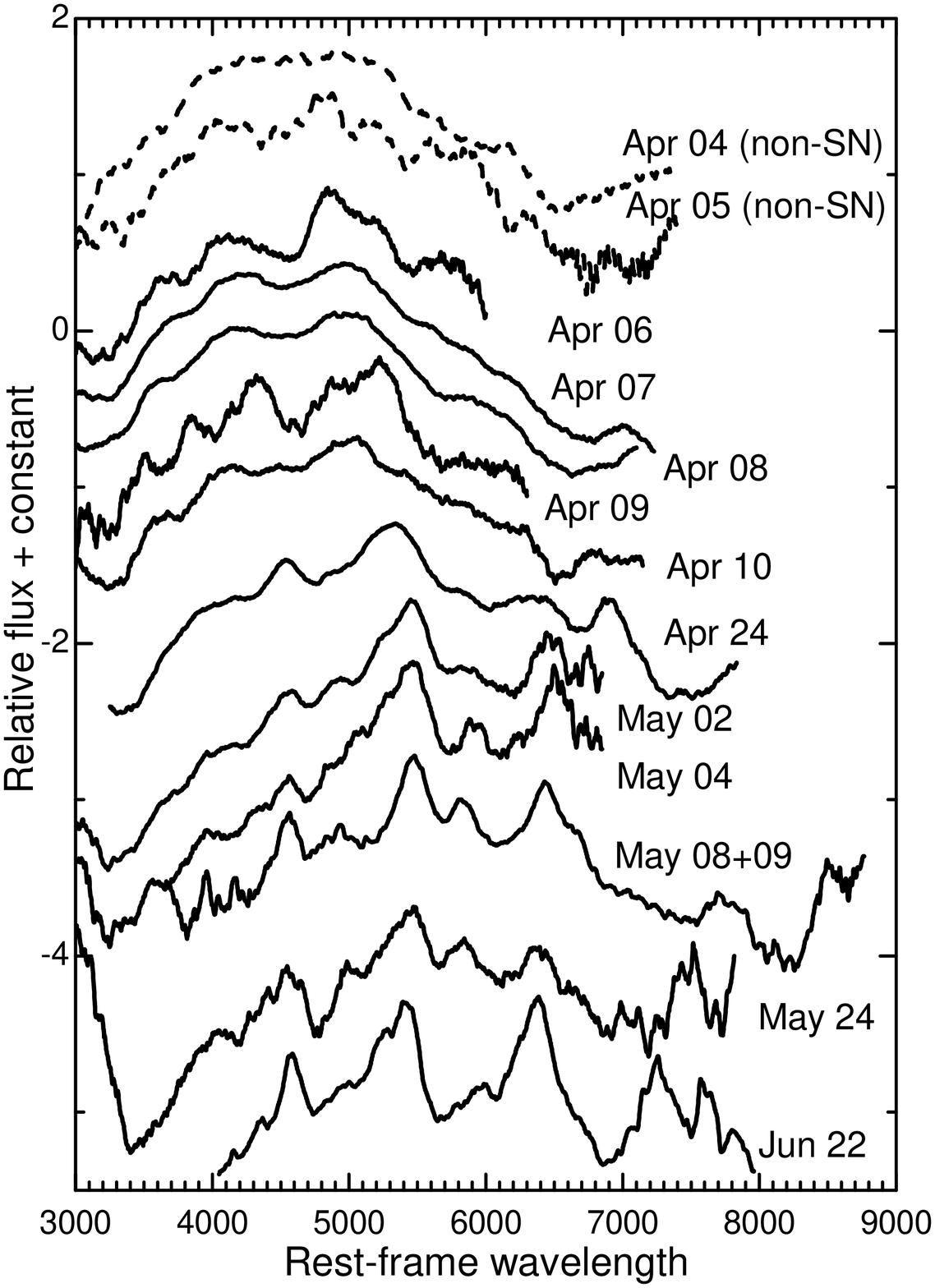]{Evolution of the derived spectra ({\em solid
lines}) in case I from April 4 to June 22. The spectra are
obtained by subtracting the April 3 OA template (except for April
4 and 7, see text) and the host galaxy template from the OT
spectra \citep{sta03b,kaw03,mat03,kos04}. Spectra shown are
smoothed for clarity. Those on April 4 and 5 ({\em dashed lines})
are not SN-like, displaying no convincing SN features.
\label{fig2}}

\figcaption[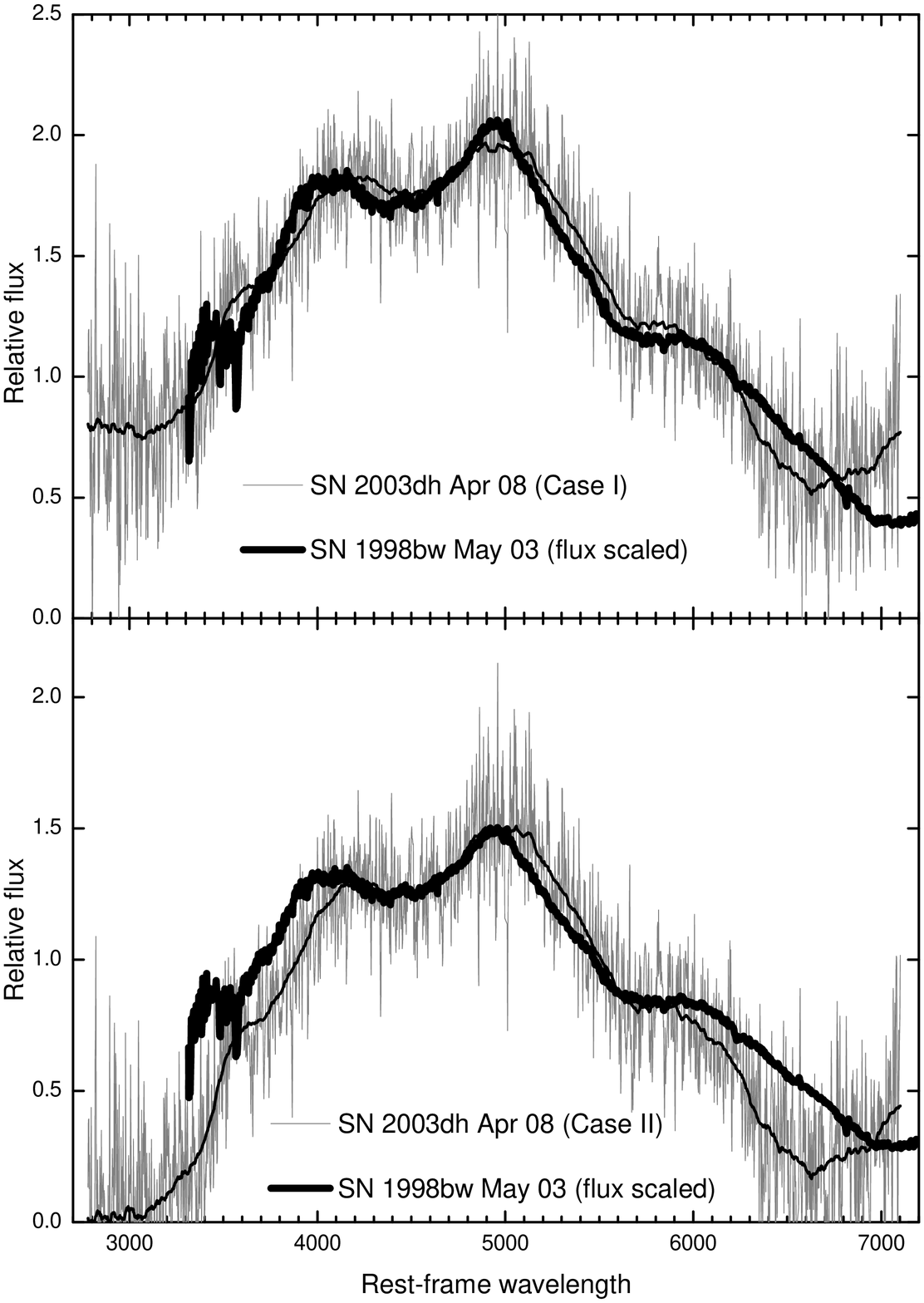]{Comparison of the derived April 8 SN 2003dh
spectrum ({\em thin gray line}: non smoothed; {\em thin black
line}: smoothed) in case I ({\em top}) and case II ({\em bottom})
with an early-time SN 1998bw spectrum ({\em thick black line};
\citealt{pat01}). \label{fig3}}

\figcaption[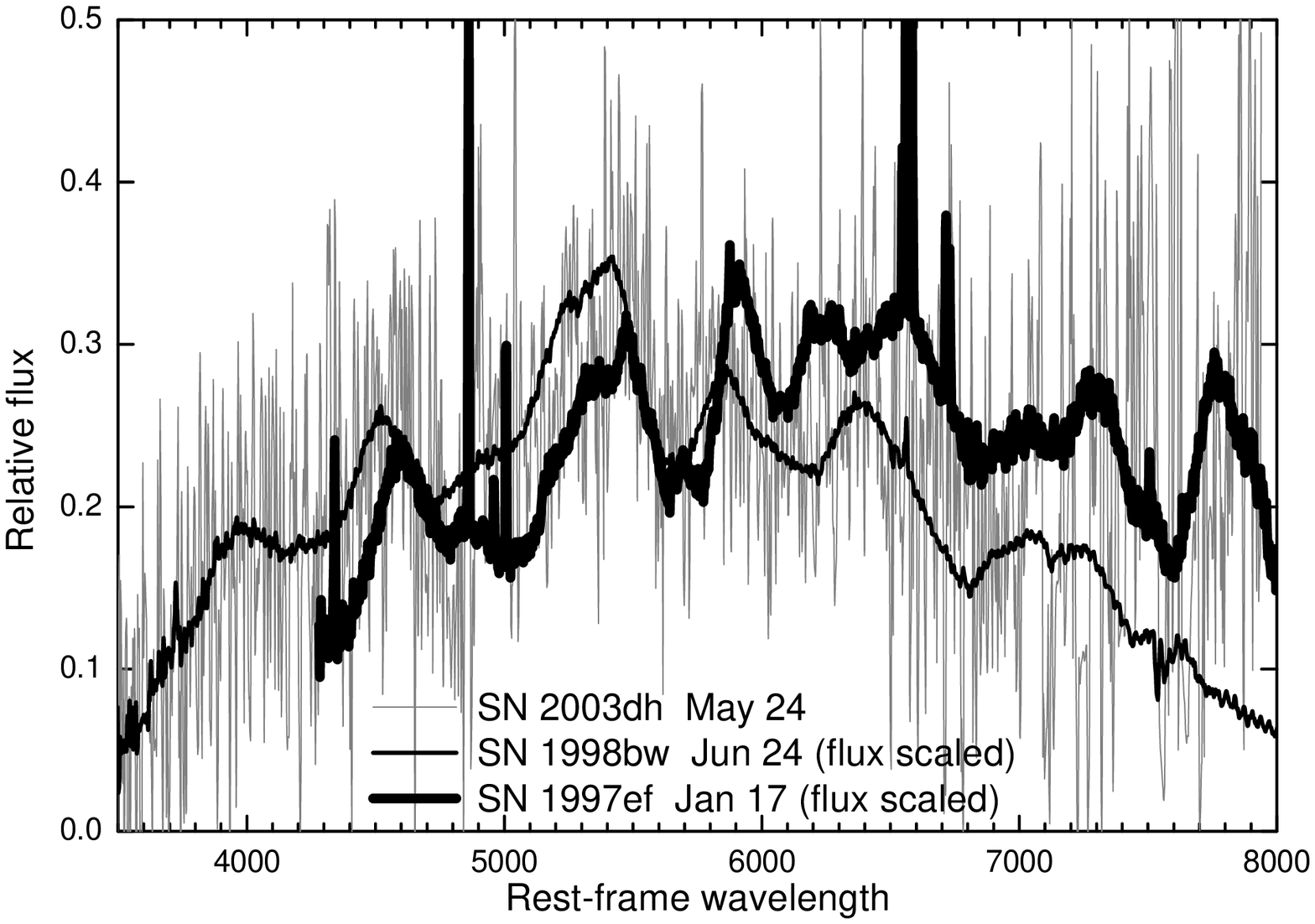]{Comparison of the derived May 24 SN 2003dh
spectrum ({\em thin gray line}) in case I with the spectra of SN
1998bw ({\em thin black line}; \citealt{pat01}) and SN 1997ef
({\em thick black line}; \citealt{mat00}) at similar epochs.
\label{fig4}}

\figcaption[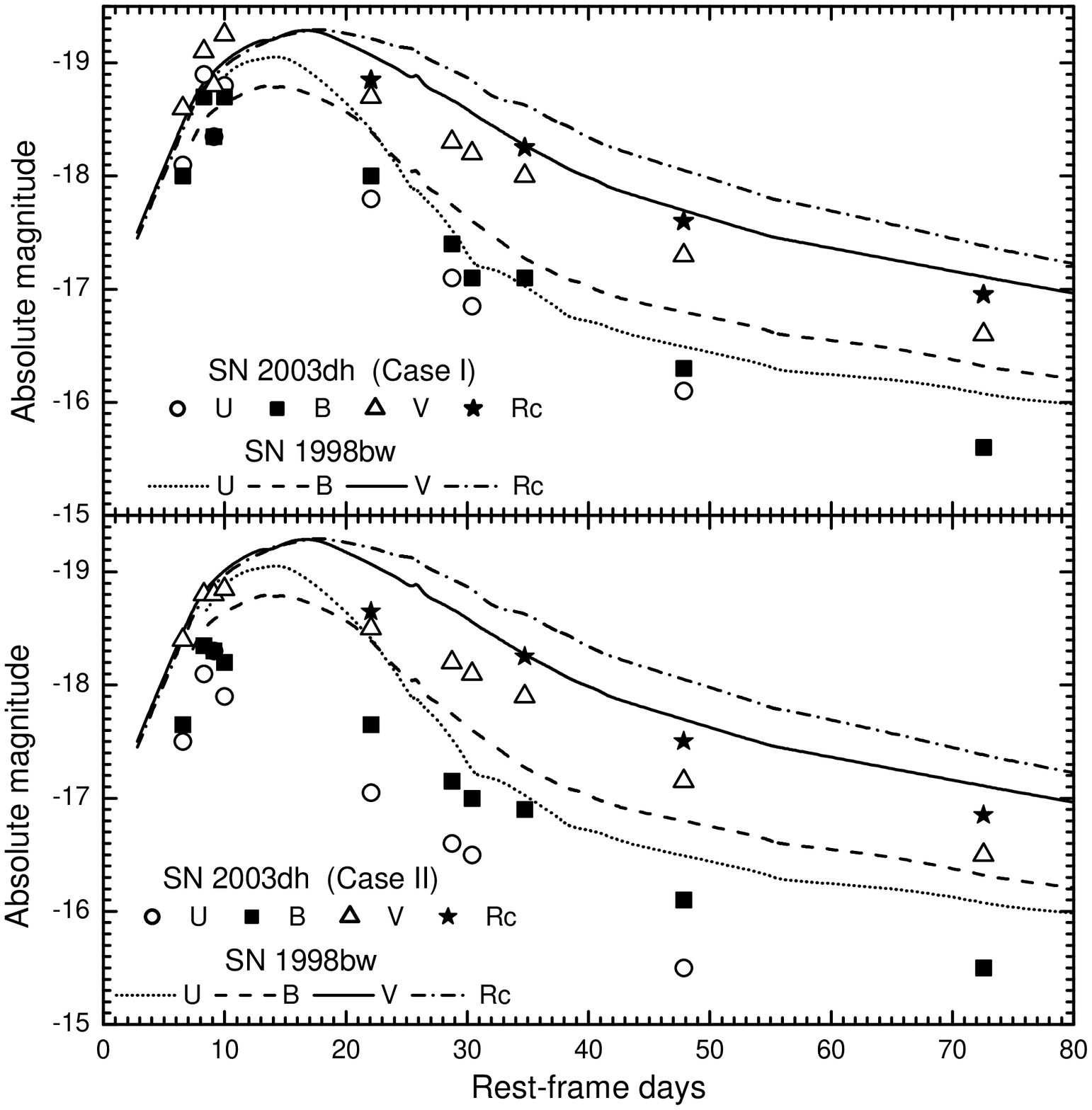]{Comparison of our rest-frame $U$ ({\em
circles}), $B$ ({\em squares}), $V$ ({\em triangles}), and $R_{\rm
C}$ ({\em stars}) LCs of SN 2003dh in case I ({\em top}) and case
II ({\em bottom}) with those of SN 1998bw ({\em dotted}, {\em
dashed}, {\em solid}, and {\em dash-dotted line}, respectively;
\citealt{gal98}). Results shown are averaged over the cases of
different OA templates. \label{fig5}}

\figcaption[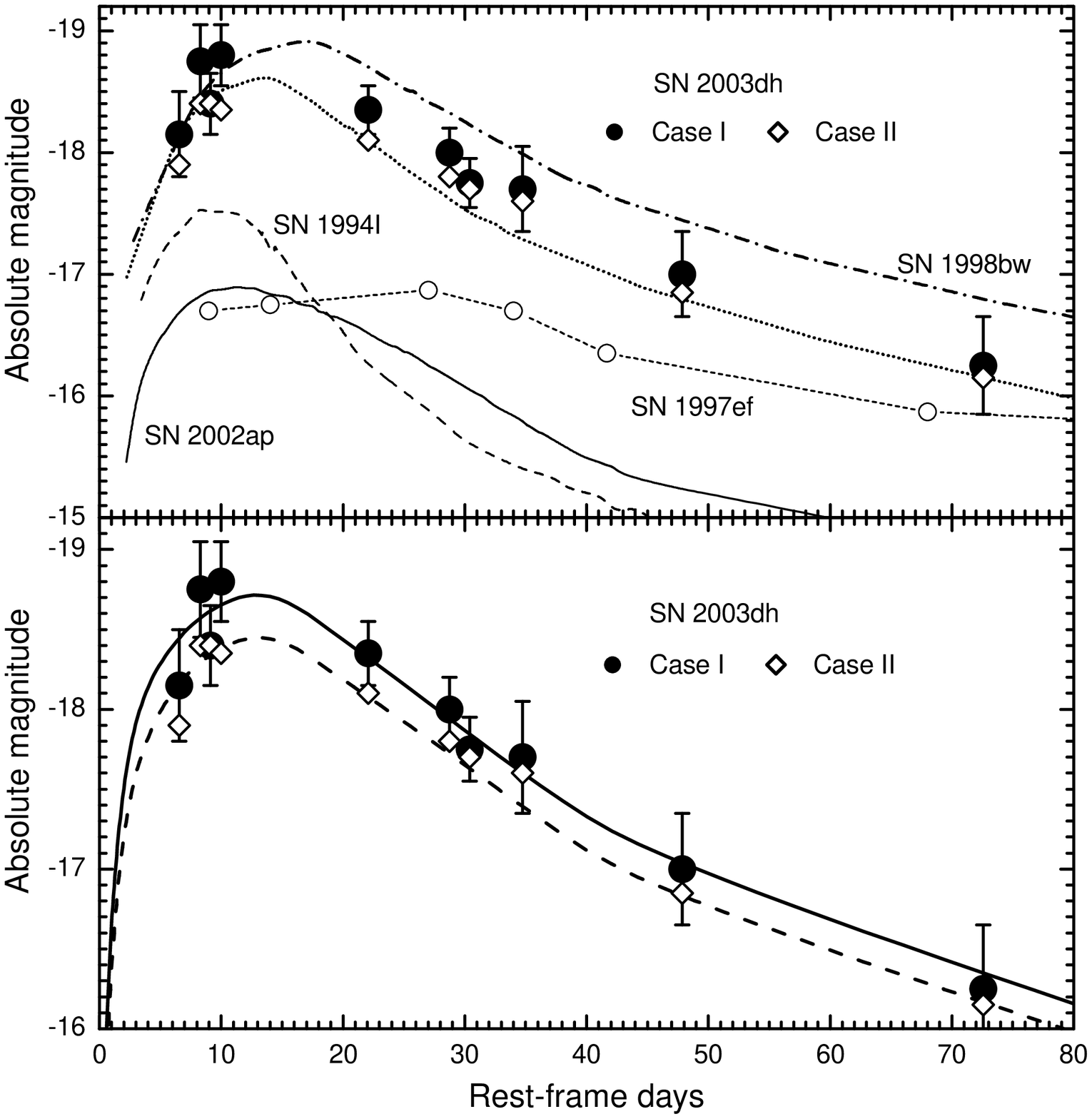]{{\em Top}: Comparison of our bolometric LCs of
SN 2003dh in case I ({\em filled circles}) and case II ({\em
diamonds}) with that suggested by (\citealt{lip04}; {\em dotted
line}) and those of (see text for references) SN 1998bw ({\em
dash-dotted line}), SN 1997ef ({\em open circles with short-dashed
line}), SN 2002ap ({\em solid line}), and SN 1994I ({\em dashed
line}). Error bars in case II are not shown for clarity. {\em
Bottom}: Comparison of our bolometric LCs of SN 2003dh with model
LCs ({\em solid line}: 7 ${\rm M}_\sun$, 3.5 $\times 10^{52}$
ergs, and 0.4 $M_\sun$ $^{56}$Ni; {\em dashed line}: 8 ${\rm
M}_\sun$, 4 $\times 10^{52}$ ergs, and 0.35 $M_\sun$ $^{56}$Ni).
\label{fig6}}

\figcaption[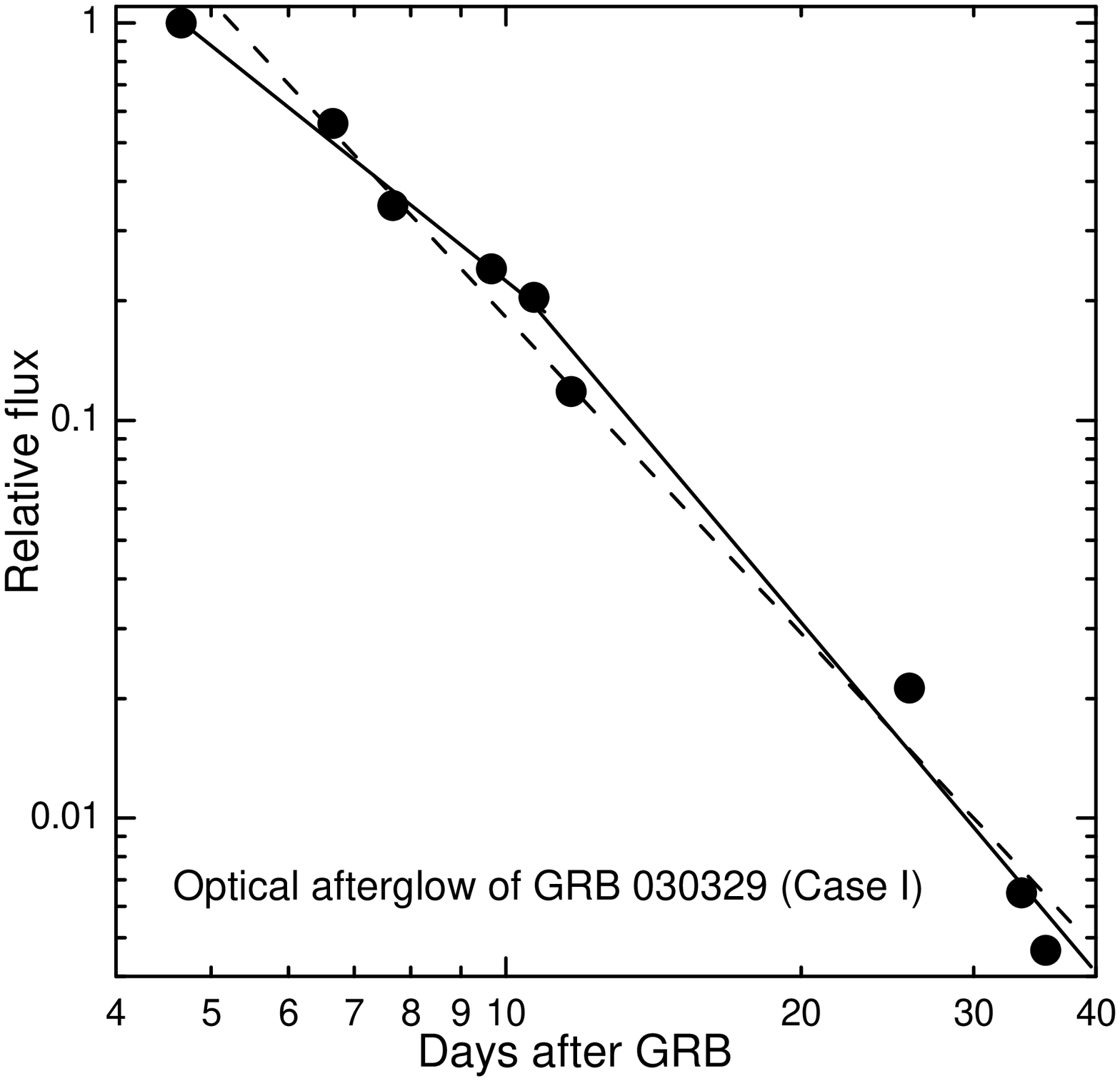]{Comparison of the OA LC of GRB 030329 after
April 3 ({\em circles}) with best-fitting single power law ({\em
dashed line}: $f\propto t^{-2.6\pm 0.1}$) and broken power law
({\em solid line}: $t\lesssim$ 10 days, $f\propto t^{-2.0\pm
0.1}$; $t\gtrsim$ 10 days, $f\propto t^{-2.9\pm 0.2}$). The LC is
derived from our spectrum decomposition in case I using the April
3 OA spectrum template, and the flux shown is normalized relative
to the template. \label{fig7}}

\begin{deluxetable}{cccccccccc}
 \tabletypesize{\scriptsize}
 \tablecaption{SN 2003dh LCs from Spectrum Decomposition (Case I)\label{table1}}
 \tablewidth{0pt}
 \tablehead{
   \colhead{Days\tablenotemark{a}}
 & \colhead{$M$($U$)\tablenotemark{b}}
 & \colhead{$M$($B$)\tablenotemark{b}}
 & \colhead{$M$($V$)\tablenotemark{b}}
 & \colhead{$M$($R_{\rm C}$)\tablenotemark{b}}
 & \colhead{BC($U$)\tablenotemark{c}}
 & \colhead{BC($B$)\tablenotemark{c}}
 & \colhead{BC($V$)\tablenotemark{c}}
 & \colhead{BC($R_{\rm C}$)\tablenotemark{c}}
 & \colhead{$M$(bol)\tablenotemark{d}}
 }
\startdata
  5.7 & -18.50 $\pm$ 0.40 & -18.40 $\pm$ 0.40 & -19.05 $\pm$ 0.35 & \nodata           &  0.00 & -0.05 & 0.30 & 0.25 & -18.60 $\pm$ 0.40\\
  6.6 & -18.10 $\pm$ 0.35 & -18.00 $\pm$ 0.35 & -18.60 $\pm$ 0.30 & \nodata           &  0.05 & -0.05 & 0.35 & 0.25 & -18.15 $\pm$ 0.35\\
  8.3 & -18.90 $\pm$ 0.25 & -18.70 $\pm$ 0.25 & -19.10 $\pm$ 0.30 & \nodata           &  0.15 &  0.00 & 0.35 & 0.30 & -18.75 $\pm$ 0.30\\
  9.1 & -18.35 $\pm$ 0.25 & -18.35 $\pm$ 0.25 & -18.80 $\pm$ 0.20 & \nodata           &  0.20 &  0.00 & 0.35 & 0.30 & -18.40 $\pm$ 0.25\\
 10.0 & -18.80 $\pm$ 0.20 & -18.70 $\pm$ 0.20 & -19.25 $\pm$ 0.20 & \nodata           &  0.20 &  0.00 & 0.35 & 0.30 & -18.80 $\pm$ 0.25\\
 22.1 & -17.80 $\pm$ 0.20 & -18.00 $\pm$ 0.20 & -18.70 $\pm$ 0.20 & -18.85 $\pm$ 0.20 & -0.30 & -0.30 & 0.35 & 0.50 & -18.35 $\pm$ 0.20\\
 28.7 & -17.10 $\pm$ 0.20 & -17.40 $\pm$ 0.20 & -18.30 $\pm$ 0.20 & \nodata           & -0.80 & -0.60 & 0.35 & 0.60 & -18.00 $\pm$ 0.20\\
 30.4 & -16.85 $\pm$ 0.20 & -17.10 $\pm$ 0.20 & -18.20 $\pm$ 0.20 & \nodata           & -0.95 & -0.60 & 0.35 & 0.60 & -17.75 $\pm$ 0.20\\
 34.7 & \nodata           & -17.10 $\pm$ 0.35 & -18.00 $\pm$ 0.35 & -18.25 $\pm$ 0.35 & -0.95 & -0.70 & 0.30 & 0.65 & -17.70 $\pm$ 0.35\\
 47.8 & -16.10 $\pm$ 0.35 & -16.30 $\pm$ 0.35 & -17.30 $\pm$ 0.35 & -17.60 $\pm$ 0.35 & -0.95 & -0.65 & 0.25 & 0.60 & -17.00 $\pm$ 0.35\\
 72.6 & \nodata           & -15.60 $\pm$ 0.35 & -16.60 $\pm$ 0.35 & -16.95 $\pm$ 0.35 & -0.70 & -0.50 & 0.30 & 0.60 & -16.25 $\pm$ 0.40\\
\enddata
\tablecomments{All magnitude data and error bars are accurate to
0.05 mag.}

 \tablenotetext{a}{Days in the rest frame ($z=0.1685$) since GRB 030329 on 2003 March 29.48 UT.}

 \tablenotetext{b}{Absolute magnitudes of derived SN spectra [$z=0.1685$, $\mu=39.54$, and $E(B-V)=0.025$].}

 \tablenotetext{c}{Bolometric corrections of SN 1998bw calculated based on \citet{gal98} and \citet{pat01}.}

 \tablenotetext{c}{Average $UVOIR$ bolometric magnitudes adopting the $B$-, $V$-, and $R_{\rm C}$-band BCs of SN 1998bw.}

\end{deluxetable}

\onecolumn

\clearpage

\plotone{f1.eps}

\plotone{f2.eps}

\plotone{f3.eps}

\plotone{f4.eps}

\plotone{f5.eps}

\plotone{f6.eps}

\plotone{f7.eps}

\end{document}